\documentclass[sigconf]{acmart}
\usepackage[ruled,vlined]{algorithm2e}

\AtBeginDocument{%
}

\renewcommand\footnotetextcopyrightpermission[1]{}
\setcopyright{none}

\setcopyright{acmlicensed}
\copyrightyear{2025}
\acmYear{2025}

\acmConference[SC '25]{International Conference for High Performance Computing, Networking, Storage, and Analysis (SC25)}{November 16--21, 2025}{St. Louis, MO}

\begin{document}

\title{Hybrid Cryptographic Monitoring System for \\ Side-Channel Attack Detection on PYNQ SoCs}

\author{Nishant Chinnasami}
\affiliation{%
  \institution{University of South Carolina}
  \city{Columbia}
  \state{SC}
  \country{USA}
}
\email{nishantc@email.sc.edu}

\author{Rasha Karakchi (Advisor)}
\affiliation{%
  \institution{University of South Carolina}
  \city{Columbia}
  \state{SC}
  \country{USA}
}
\email{karakchi@cec.sc.edu}

\begin{abstract}
AES-128 encryption is theoretically secure but vulnerable in practical deployments due to timing and fault injection attacks on embedded systems. This work presents a lightweight dual-detection framework combining statistical thresholding and machine learning (ML) for real-time anomaly detection. By simulating anomalies via delays and ciphertext corruption, we collect timing and data features to evaluate two strategies: (1) a statistical threshold method based on execution time and (2) a Random Forest classifier trained on block-level anomalies. Implemented on CPU and FPGA (PYNQ-Z1), our results show that the ML approach outperforms static thresholds in accuracy, while maintaining real-time feasibility on embedded platforms. The framework operates without modifying AES internals or relying on hardware performance counters. This makes it especially suitable for low-power, resource-constrained systems where detection accuracy and computational efficiency must be balanced.
\end{abstract}


\maketitle
\section{Introduction}
The Advanced Encryption Standard (AES) has become the cornerstone for securing data across diverse platforms, from cloud data centers to embedded IoT devices \cite{karakchi2025toward}. Among its variants, AES-128 remains widely adopted for its optimal balance between performance and security. However, despite its cryptographic soundness, real-world implementations often leak unintended information on the side channel \cite{kocher1996timing, bernstein2005cache}, opening the door to physical attacks or timing-based attacks.

Timing attacks exploit variations in encryption latency to infer secret keys. Kocher's seminal work~\cite{kocher1996timing} exposed how control flow and cache behavior can leak timing differences. Bernstein extended this by showing that even remote attacks could succeed on shared systems \cite{bernstein2005cache}, highlighting how software-only observations can defeat theoretically secure algorithms. Fault injection attacks, such as differential fault analysis (DFA) \cite{biham1997differential}, manipulate the computation process, potentially revealing internal state values or leading to predictable ciphertext corruption.

Countermeasures such as constant-time coding and masking~\cite{cryptoeprint:2017:106, choi2020advanced} mitigate some risks but often require hardware modifications or introduce runtime overhead, which is infeasible in low-power or embedded systems. Some defenses rely on hardware duplication or error-detection codes, but these are impractical for widespread deployment in low-cost systems.

Recently, machine learning-based anomaly detection has gained traction, analyzing traces like power or execution time to detect malicious behavior~\cite{wang2021machine, yu2023ml, zhang2020random, abdellatif2021ml, liu2023deep}. Although promising, such approaches typically require large labeled datasets and computationally heavy models, making real-time deployment on embedded devices difficult.

\begin{figure}[ht]
\centering
\includegraphics[width=0.5\textwidth]{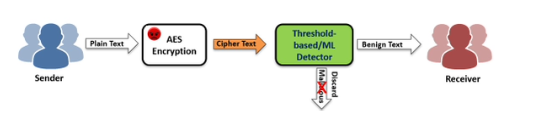} 
\caption{Proposed Detection Framework}
\label{fig:system}
\end{figure}

In this work, we develop a lightweight threshold-based timing anomaly detector Figure \ref{fig:system} that flags encryption anomalies using only execution time. This method is non-intrusive, requires no access to internal AES logic, and avoids external sensors, making it ideal for embedded deployments. To further enhance granularity, we complement this with a supervised machine learning approach that uses timing and ciphertext features.

We evaluated both methods on a multicore CPU and the Xilinx PYNQ-Z1 FPGA SoC~\cite{xilinxpynqz1}, comparing performance, accuracy, and feasibility. Our results provide insight into how statistical and ML-based detectors behave under controlled anomaly injection and serve as a foundation for future hybrid cryptographic monitoring systems.

\section{Methodology}
In this work, we develop and evaluate a statistical timing-based anomaly detection framework to enhance the security of AES-128 encryption. Our methodology consists of anomaly injection, parallel encryption execution, timing measurement, and threshold-based anomaly detection. We evaluated this methodology on both a standard multi-core CPU (8 logical cores and 3.2 GHz) environment and a Xilinx PYNQ Z1 platform, which is equipped with a dual-core ARM Cortex-A9 processor.

\subsection{Anomaly Injection Mechanism}
To simulate potential threats, we inject two types of anomalies into selected encryption blocks:

\begin{itemize}
\item \textbf{Timing Delay Injection:} A short delay (5–20 ms) is added before encryption to simulate cache contention or intentional software slowdowns, mimicking timing-based side-channel behavior.
\item \textbf{Fault Injection (Bit Flip):} The first byte of the plaintext block is XORed with \texttt{0xFF} before encryption, simulating data corruption from voltage glitches or electromagnetic interference.
\end{itemize}

Each anomaly-injected block is labeled malicious. The injection ratio $r$ determines the percentage of blocks modified (typically 10\%–30\%). These modifications ensure that both latency- and content-based deviations are present in the data set. The complete procedure is described in Algorithm~\ref{alg:anomaly-detect}.

\begin{algorithm}[htbp]
\SetAlgoLined
\footnotesize
\KwIn{Number of blocks $n$, CPU cores $c$, Anomaly ratio $r$}
\KwOut{Encrypted blocks, anomaly detection report}

Generate $n$ random plaintext blocks\;
For each block:
With probability $r$, inject anomaly (delay or fault)\;

Parallelize using $c$ cores:
For each block:
\If{anomaly == delay}{
Sleep for a short random time\;
}
\If{anomaly == fault}{
Flip first byte of block\;
}
Pad/truncate block to 16 bytes\;
Encrypt block with AES-128 in ECB mode\;
Record encryption time\;

Compute time threshold for anomaly detection\;
For each result:
\If{encryption\_time > threshold}{
Mark as malicious\;
}

Compute detection metrics (accuracy, false positives, etc.)\;
Save results to Excel and display sample outputs\;

\caption{AES Encryption and Timing-Based Anomaly Detection}
\label{alg:anomaly-detect}
\end{algorithm}

AES-128 encryption is implemented in Electronic Code Book (ECB) mode using the PyCryptodome library \cite{pycryptodome}. Each plaintext input is a 16-byte randomly generated block. We deploy the encryption framework on the PYNQ Z1 platform, which enables parallel processing using Python's \texttt{multiprocessing} module. This setup mimics real-time cryptographic execution under multi-core embedded environments.

\subsection{Timing-Based Anomaly Detection}
The threshold-based detection uses the following statistical rule:

\[
T = \mu + 3 \times \frac{\max - \min}{n}
\]

where \(\mu\) is the mean encryption time, \(\max\) and \(\min\) are the observed maximum and minimum times, and \(n\) is the total number of blocks. This rule is inspired by the empirical "3-sigma" rule used in statistical outlier detection~\cite{hand2001idiot}. Any block whose execution time exceeds \(T\) is flagged as potentially malicious.
\subsection{Machine Learning-Based Detection}

To overcome the limitations of timing-only detection, we implement a supervised machine learning method using a Random Forest classifier. Features extracted from each block include encryption time and ciphertext bytes. The classifier is trained to detect anomalies such as fault-injected or delay-affected blocks. Algorithm \ref{alg:ml-detection} describes the ML-based anomaly detection procedure.

\begin{algorithm}[htbp]
\SetAlgoLined
\footnotesize
\KwIn{$N$: number of plaintext blocks, $p$: malicious percentage, $c$: CPU cores}
\KwOut{Encrypted blocks, ML detection report}

\textbf{Step 1: Generate Blocks} \\
\For{$i \leftarrow 1$ \KwTo $N$}{
  Generate random 16-byte block $B_i$\;
  Inject anomaly with probability $p/100$\;
}

\textbf{Step 2: Encrypt in Parallel} \\
\ForEach{($B_i$, $i$, anomaly) \textbf{in parallel with} $c$ cores}{
  \If{anomaly}{
    Randomly select anomaly type $\in$ \{delay, fault\}\;
    \If{delay}{Sleep for short random time\;}
    \ElseIf{fault}{Flip first byte of $B_i$\;}
  }
  Pad/trim $B_i$ to 16 bytes\;
  Encrypt $B_i$ using AES-128-ECB\;
  Record encryption time $t_i$\;
}

\textbf{Step 3: Extract Features} \\
Create dataset with timing $t_i$ and original block bytes\;

\textbf{Step 4: Train ML Classifier} \\
Split dataset into train/test sets\;
Train Random Forest classifier\;
Predict labels on test set\;

\textbf{Step 5: Evaluate and Report} \\
Calculate TP, FP, FN, Accuracy\;
Compute threshold $T = \mu + 3 \times \frac{\max - \min}{N}$\;
Save results to Excel\;

\caption{AES Encryption with ML-Based Anomaly Detection}
\label{alg:ml-detection}
\end{algorithm}

\subsection{Experimental Setup}
The experiments were carried out on two platforms:

\begin{itemize}
\item \textbf{CPU Platform:} A multi-core desktop with Intel i7 CPU and 16GB RAM, using Python’s multiprocessing for parallel encryption and timing.
\item \textbf{Embedded Platform:} Xilinx PYNQ-Z1 featuring a dual-core ARM Cortex-A9 CPU running lightweight Linux. All logic was implemented in native Python using onboard resources.
\end{itemize}

In both setups, timing logs, anomaly labels, and output metrics were stored for offline analysis. This ensured a consistent dataset for both threshold and ML-based evaluation.

\section{Deployment and Evaluation}

We evaluated both detection methods on a multi-core CPU and the PYNQ-Z1 FPGA, representing general-purpose and embedded environments.

The threshold method offered fast, low-overhead detection but missed subtle anomalies. The Random Forest classifier, using timing and ciphertext features, improved accuracy with inference latency under 5\,ms on the PYNQ-Z1.

Both approaches performed consistently across platforms. The ML model used under 30\% of FPGA resources, confirming its suitability for embedded use. Thresholding ensures speed and simplicity, while ML enhances detection in complex scenarios. Future work includes integrating side-channel features and tighter hardware coupling.


\begin{acks}
This work was supported by the Office of Undergraduate Research and the McNair Junior Fellowship at the University of South Carolina. The authors thank Rye Stahle-Smith for his assistance with hardware testing and experimental setup.
\end{acks}

\bibliographystyle{unsrt}
\bibliography{main}

\begin{thebibliography}{10}

\bibitem{karakchi2025toward}
Rasha Karakchi, Rye Stahle-Smith, Nishant Chinnasami, and Tiffany Yu.
\newblock Toward a lightweight, scalable, and parallel secure encryption engine.
\newblock {\em arXiv preprint arXiv:2506.15070}, 2025.

\bibitem{kocher1996timing}
Paul~C. Kocher.
\newblock Timing attacks on implementations of diffie-hellman, rsa, dss, and other systems.
\newblock In {\em CRYPTO}, 1996.

\bibitem{bernstein2005cache}
Daniel~J. Bernstein.
\newblock Cache-timing attacks on aes.
\newblock \url{http://cr.yp.to/antiforgery/cachetiming-20050414.pdf}, 2005.
\newblock Accessed: 2025-06-22.

\bibitem{biham1997differential}
Eli Biham and Adi Shamir.
\newblock Differential fault analysis of secret key cryptosystems.
\newblock In Burton S.~Kaliski Jr., editor, {\em Advances in Cryptology — CRYPTO ’97}, volume 1294 of {\em Lecture Notes in Computer Science}, pages 513--525. Springer, 1997.

\bibitem{cryptoeprint:2017:106}
Daniel Genkin, Lev Pachmanov, Itamar Pipman, Eran Tromer, and Yuval Yarom.
\newblock Ecdsa key extraction from mobile devices via nonintrusive physical side channels.
\newblock In {\em Proceedings of the 2016 ACM SIGSAC conference on computer and communications security}, pages 1626--1638, 2016.

\bibitem{choi2020advanced}
Subidh Ali, Xiaofei Guo, Ramesh Karri, and Debdeep Mukhopadhyay.
\newblock Fault attacks on aes and their countermeasures.
\newblock pages 163--208, 2016.

\bibitem{wang2021machine}
Alshaibi~Ahmed Jamal, Al-Ani~Mustafa Majid, Anton Konev, Tatiana Kosachenko, and Alexander Shelupanov.
\newblock A review on security analysis of cyber physical systems using machine learning.
\newblock {\em Materials today: proceedings}, 80:2302--2306, 2023.

\bibitem{yu2023ml}
Weiwei Shan, Shuai Zhang, and Yukun He.
\newblock Machine learning based side-channel-attack countermeasure with hamming-distance redistribution and its application on advanced encryption standard.
\newblock {\em Electronics Letters}, 53(14):926--928, 2017.

\bibitem{zhang2020random}
Avneet Gupta and RajBala Simon.
\newblock Enhancing security in cloud computing with anomaly detection using random forest.
\newblock pages 1--6, 2024.

\bibitem{abdellatif2021ml}
Lars Bauer, Hassan Nassar, Nadir Khan, J{\"u}rgen Becker, and J{\"o}rg Henkel.
\newblock Machine-learning-based side-channel attack detection for fpga socs.
\newblock {\em IEEE Transactions on Circuits and Systems for Artificial Intelligence}, 2024.

\bibitem{liu2023deep}
Lo{\"\i}c Masure, C{\'e}cile Dumas, and Emmanuel Prouff.
\newblock A comprehensive study of deep learning for side-channel analysis.
\newblock {\em IACR Transactions on Cryptographic Hardware and Embedded Systems}, pages 348--375, 2020.

\bibitem{xilinxpynqz1}
Xilinx Inc.
\newblock Pynq-z1: Python productivity for zynq.
\newblock \url{https://www.pynq.io/board.html}, 2018.
\newblock Accessed: 2025-06-22.

\bibitem{pycryptodome}
Dario Legrandin.
\newblock Pycryptodome: Python cryptographic library.
\newblock \url{https://www.pycryptodome.org/}, 2018.
\newblock Version accessed: 2025.

\bibitem{hand2001idiot}
D.~J. Hand and K.~Yu.
\newblock Idiot’s bayes—not so stupid after all?
\newblock {\em International Statistical Review}, 69(3):385--398, 2001.

\end{thebibliography}
\end{document}